\documentclass[aps,prb,twocolumn,showpacs,amsmath,amssymb,superscriptaddress]{revtex4-1}

\usepackage[utf8]{inputenc}

\usepackage{graphicx}
\usepackage[colorlinks=true,citecolor=blue]{hyperref}
\usepackage{units}

\usepackage{verbatim} 

\renewcommand{\vec}[1]{\mathbf{#1}}

\def\kB {k_\text{B}}

\begin{document}

\title{Rectification of thermal fluctuations in a chaotic cavity heat engine}

\author{Bj\"orn Sothmann}
\affiliation{D\'epartement de Physique Th\'eorique, Universit\'e de Gen\`eve, CH-1211 Gen\`eve 4, Switzerland}
\author{Rafael S\'anchez}
\affiliation{Instituto de Ciencia de Materiales de Madrid (ICMM-CSIC), Cantoblanco, E-28049 Madrid, Spain}
\author{Andrew N. Jordan}
\affiliation{Department of Physics and Astronomy, University of Rochester, Rochester, New York 14627, USA}
\author{Markus B\"uttiker}
\affiliation{D\'epartement de Physique Th\'eorique, Universit\'e de Gen\`eve, CH-1211 Gen\`eve 4, Switzerland}

\date{\today}

\begin{abstract}
We investigate the rectification of thermal fluctuations in a mesoscopic on-chip heat engine. The engine consists 
of a hot chaotic cavity capacitively coupled to a cold cavity which rectifies the excess noise and generates a directed current. 
The fluctuation-induced directed current depends on the energy asymmetry of the transmissions of the contacts of the cold cavity to the leads and is proportional to the temperature difference. We discuss the channel dependence of the maximal power output of the heat engine and its efficiency.
\end{abstract}


\maketitle
\section{\label{sec:introduction}Introduction}
Rectification is central to the operation of electrical circuits. More than 60 years ago, Leon Brillouin, then at IBM, raised the question of whether an electric circuit consisting of a resistor and a diode can become a Maxwell demon rectifying its own thermal fluctuations.~\cite{brillouin_can_1950} Using inappropriate generalizations of Langevin dynamics for systems with nonlinear diffusion coefficients could indeed lead to such rectification, in obvious violation of the second law of thermodynamics.~\cite{marek_note_1959}
Brillouin's paradox was solved by taking into account the diode's contact potentials.~\cite{van_kampen_non-linear_1960} Later, it was shown that a system with diode and resistor at two different temperatures cannot exceed Carnot efficiency~\cite{sokolov_energetics_1998} in agreement with the second law.

Nowadays, thermoelectrics is of increasing importance.  In the continuing quest for smaller scale electric circuits the evacuation of heat proves to be a major obstacle. Therefore it is interesting to explore whether some of the energy that is dissipated can be harvested  and put to use. 

There are many ways to generate directed currents. In recent years Brownian particles in ratchets subject to periodic driving have found much interest in very different fields of science.~\cite{astumian_brownian_2002} Here we are concerned with a more subtle form of driving: The only external agent acting on the system is noise that can be generated by an external thermal equilibrium bath. External noise can generate directed currents even in periodic potentials with inversion symmetry if the noise power depends on the location of the Brownian particle.~\cite{buettiker_transport_1987,van_kampen_relative_1988,blanter_rectification_1998,olbrich_ratchet_2009,olbrich_classical_2011} Such state dependent diffusion is also at the origin of the difficulties encountered by Brillouin.~\cite{brillouin_can_1950}

\begin{figure}
	\includegraphics[width=\columnwidth]{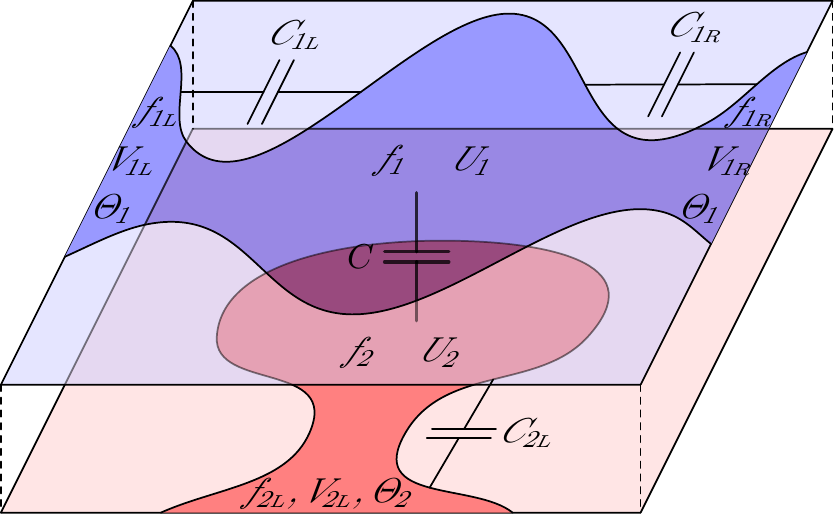}
	\includegraphics[width=\columnwidth]{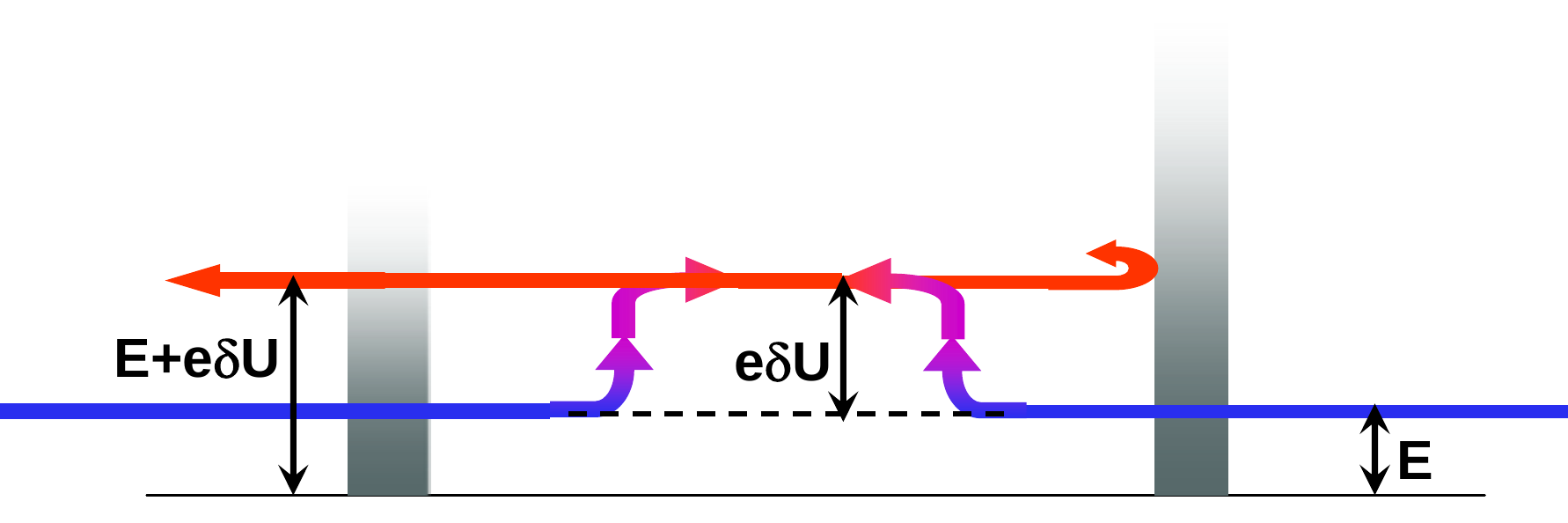}
	\caption{\label{fig:model}(Color online) Schematic of the double cavity and operation scheme: Electrons enter the cavity, gain an energy $e\delta U$ and then leave the cavity again, experiencing different transmissions of the contacts.}
\end{figure}

In this paper, we investigate the rectification of thermal fluctuations into a directed electric current in a mesoscopic heat engine. The latter consists of two capacitively coupled chaotic cavities arranged in a three-terminal geometry as shown Fig.~\ref{fig:model}. The upper cavity is the rectifier and is connected via two contacts with energy-dependent transmissions to electron reservoirs. The lower cavity provides the external source of thermal noise. It is connected via only a single contact to another electron reservoir. The energy dependence of the contact transmissions is a generic feature of mesoscopic conductors and leads to an intrinsic nonlinearity in the upper cavity.~\cite{jordan_gap_2008}
The three-terminal setup allows for separated heat and charge currents in contrast to two-terminal setups where these currents are necessarily aligned.

The mechanism giving rise to the current is shown schematically in Fig.~\ref{fig:model}. Electrons enter the cold upper cavity at an energy $E$. They absorb the energy $e\delta U$ from the fluctuating potential generated by the hot lower cavity in the upper cavity and afterwards leave the cold cavity again. As the transmissions through the upper cavity's contacts are energy-dependent, the ratio between transport processes involving the left and right lead is different at energies $E$ and $E+e\delta U$, thus giving rise to a net electrical current through the cavity. The rectification is controlled by Coulomb coupling which is dominant at low temperatures, whereas at higher temperatures phonon effects become important.~\cite{entin-wohlman_three-terminal_2010}

Thermoelectric properties of both open mesoscopic cavities  ~\cite{godijn_thermopower_1999,llaguno_observation_2003} and in the Coulomb blockade limit~\cite{molenkamp_sawtooth-like_1994,staring_coulomb-blockade_1993,dzurak_thermoelectric_1997} have been of interest. The quantization of energy levels in small dots leads to exceptional thermo-electrical properties.~\cite{edwards_cryogenic_1995} Both in two- and three-terminal structures the limit in which the ratio of electric to heat current is given by the ratio of the charge to an energy quantum can be reached.~\cite{humphrey_reversible_2002,sanchez_optimal_2011} Still, while the efficiency of such nanoengines can in principle be optimal leading to an infinite figure of merit $ZT$\cite{mahan_best_1996}, the current they deliver is small, typically of the order of $\unit[1]{pA}$. It is therefore of interest to explore how power output and efficiency scale as dots are opened and turned into cavities with contacts that permit currents that are much larger than the tunneling current of a Coulomb-blockaded quantum dot.

The physics of Coulomb coupled conductors is of interest in nanophysics for on-chip charge detectors,~\cite{jordan_qubit_2006} quantum Hall edge states,~\cite{le_sueur_energy_2010}  and the Coulomb drag in which one system that carries a current induces a current in a nearby unbiased conductor.~\cite{laroche_positive_2011,mortensen_coulomb_2001,levchenko_coulomb_2008,sanchez_mesoscopic_2010} In the setup of Fig.~\ref{fig:model} the current carrying conductor in the Coulomb drag problem is replaced by an unbiased but hot conductor. 

Our paper is organized as follows: In Sec.~\ref{sec:model}, we present our model of the double cavity and describe our theoretical approach. Our results are presented in Sec.~\ref{sec:results} and conclusions are drawn in Sec.~\ref{sec:conclusions}. Calculational details are presented in the Appendices.

\section{\label{sec:model}Model and method}
We investigate transport through two capacitively coupled open quantum dots with mutual capacitance $C$, cf. Fig~\ref{fig:model}. Each cavity $i=1,2$ is coupled via quantum point contacts (QPCs) to electronic reservoirs $r=\text{L},\text{R}$. The latter ones are assumed to be in local equilibrium and described by a Fermi distribution $f_{ir}(x)=\{\exp[(x-\mu_{ir})/\kB \Theta_{i}]+1\}^{-1}$ with temperature $\Theta_{i}$ and chemical potential $\mu_{ir}$. Interaction effects are captured by capacitive couplings $C_{ir}$ between the cavities and their respective reservoirs that leads to screening of the potential fluctuations. 

We consider the system in the semiclassical limit where the number $N_{ir}$ of open transport channels~\cite{van_wees_quantum_1991,patel_properties_1991,roessler_transport_2011} in the QPCs is large under conditions at which dephasing destroys phase information but preserves energy. 
We can, thus, characterize the chaotic cavities by a distribution function $f_i(E)$ that depends on energy only, and focus on a semi-classical description of the physics without coherence.~\cite{blanter_semiclassical_2000} For later convenience, we write the distribution function as 
\begin{equation}
	f_i=\frac{\sum_r T_{ir}f_{ir}}{\sum_r T_{ir}}+\delta f_i.
\end{equation}
Here, the first term describes the average value of $f_i$ and is given by the average of the distributions of the reservoirs weighted with the transmission $T_{ir}$ of the respective QPC. The second term $\delta f_i$ describes fluctuations of $f_i$ around its average. Additionally, each cavity is characterized by its potential $U_i$ which also fluctuates by $\delta U_i$.

We assume the transmissions to be energy dependent which we model to first order as $T_{ir}=T_{ir}^0-eT'_{ir}\delta U_i$. The energy dependent transmission leads to a nonlinear current voltage characteristic. Even without external noise such a nonlinearity requires a self-consistent treatment.~\cite{buettiker_capacitance_1993,christen_gauge-invariant_1996} In our system,  Fig.~\ref{fig:model}, we need a self-consistent treatment not only of the average Hartree potential but in addition the fluctuating potentials. 
We remark that while the energy-independent part $T^0_{ir}$ scales linearly with the number of open transport channels $N_{ir}$, the energy-dependent part $T'_{ir}$ is independent of $N_{ir}$.

The starting point of our theoretical investigation is a kinetic equation for the distribution functions $f_i$ (see, e.g., Ref.~\onlinecite{nagaev_frequency_2004}),
\begin{equation}\label{eq:kinetic}
	e\nu_{i\text{F}}\frac{df_i}{dt}=e\nu_{i\text{F}}\frac{\partial f_i}{\partial U_i}\dot U_i+\frac{e}{h}\sum_r T_{ir}(f_{ir}-f_i)+\delta i_\Sigma,
\end{equation}
where $\nu_{i\text{F}}$ denotes the density of states of cavity $i$. It describes the change of charge in a given energy interval due to changes in the potential $U_i$, in- and outgoing electron currents through the QPCs as well as their fluctuations $\delta i_\Sigma$ where the index $\Sigma$ indicates summation over all contacts $r$ of cavity $i$.

Expressing the charge inside the cavities via the distribution functions $f_i$ as well as via the capacitances and potentials, we obtain a relation between $\delta f_i$ and $\delta U_i$ which allows us to transform the kinetic equation~\eqref{eq:kinetic} into a Langevin equation for $\delta U_i$. Neglecting terms that are cubic and higher in the potential fluctuations, the latter can be converted into a nonlinear Fokker-Planck equation with a diffusion function that depends on the cavity potential. The nonlinearity of the Langevin equation leads to subtleties in the interpretation of the stochastic integral (known as the Itô and Stratonovich problem) that gives rise to different Fokker-Planck equations.   The ``kinetic prescription'' of Klimontovich~\cite{klimontovich_ito_1990} provides a steady-state solution of Eq.~\eqref{eq:kinetic} that is in global thermal equilibrium, thus avoiding the Brillouin paradox mentioned in the introduction.

We stress that the nonlinearity is the technical origin of rectification in the cavity. A linear system would not exhibit this feature. From the Klimontovich-Fokker-Planck equation we obtain $\langle\delta U_i\rangle$ and $\langle\delta U_i\delta U_j\rangle$, cf. Appendixes~\ref{app:kinetic}-\ref{app:currents} for details.  The charge currents between upper cavity and contact $r$ are given by
\begin{equation}
	I_{1r}=\frac{e}{h}\int dE T_{1r}(f_{1r}-f_{1})+\delta I_r.
\end{equation}

\section{\label{sec:results}Results}
The critical nonlinearity of the cavity is quantified by the amount of symmetry-breaking in the energy-derivatives of the transmissions of the upper cavity, given by the rectification parameter $\Lambda$:
\begin{equation}
	\Lambda =\frac{G'_{1\text{L}}G_{1\text{R}}-G'_{1\text{R}}G_{1\text{L}}}{G_{1\Sigma}^2},
\end{equation}
where $G_{ir}=(e^2/h) T^0_{ir}$ and $G'_{ir}=(e^3/h) T'_{ir}$.    We will see below that it is the rectification conversion factor between energy and charge for an unbiased cavity.  This parameter appears in many places in terms of interaction corrections.  For example, if we consider the uncoupled upper cavity ($C=0$, $i=1$), the single cavity conductance $G_1 = G_{1L} G_{1R}/G_{1\Sigma}$ (which is the series combination of the left and right leads) has an interaction correction of $\Delta G_1 = - 2 C_{1\mu} G_{\Sigma} \Lambda^2 k_B \Theta_1/C_{1\Sigma}^2$, where $C_{1\Sigma}=\sum_r C_{1r}$ is the total capacitance of the upper cavity and $C_{1\mu}^{-1} = (e^2\nu_{1\text{F}})^{-1} +  C_{1\Sigma}^{-1}$ is its electrochemical capacitance.  We note that while $G_1$ scales with the channel number $N$, the correction $\Delta G_1$
is of order $\propto N^{-1}$ (quantum corrections in a coherent cavity~\cite{kupferschmidt_temperature_2008} are of order $\propto N^0$). The rectification parameter $\Lambda$ also appears in the second-order conductance.  

\begin{figure}
        \includegraphics[width=\columnwidth]{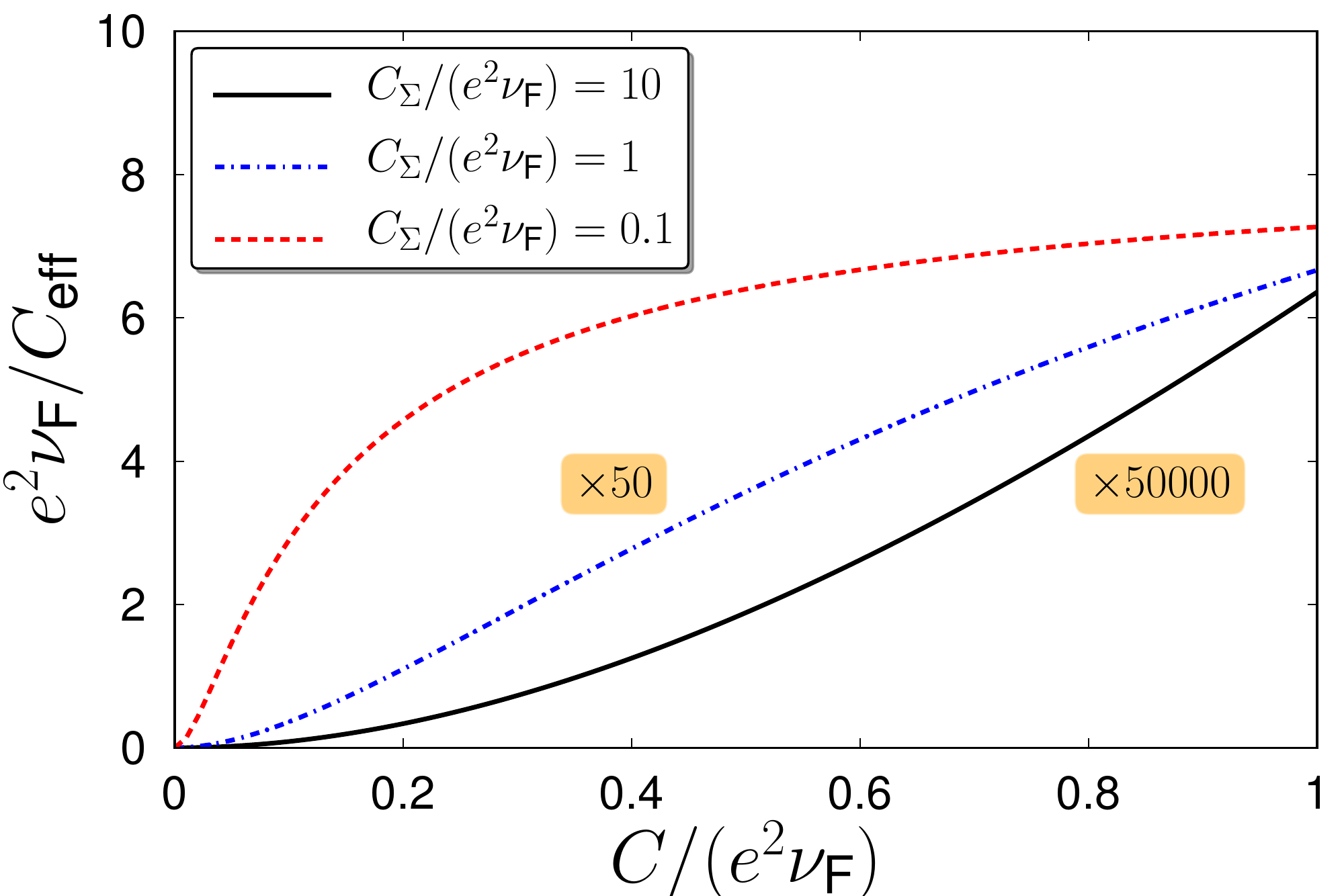}
	\caption{\label{fig:Ceff}Inverse effective capacitance as a function of the capacitance between the two cavities $C$ for different values of $e^2\nu_{\text{F}}/C_{\Sigma}$. Note the different scale for the various curves.}
\end{figure}

We now turn to the coupled cavities, cf. Fig.~1. Applying different temperatures $\Theta_1$ and $\Theta_2$ to the reservoirs that couple to cavity 1 and 2, respectively, while keeping the electrical contacts grounded ($V_{ir}=0$), we find to leading order in the nonlinearity a charge current through the cavity given by
\begin{equation}\label{eq:current}
	\langle I_\text{1L}\rangle=\frac{\Lambda}{\tau_{RC}} \kB(\Theta_1-\Theta_2),
\end{equation}
where we assumed identical capacitances and densities of states for the two cavities. Now, we give a physical interpretation of each term in Eq.~\eqref{eq:current}, the rectified current. 
$\tau_{RC}=C_\text{eff}/G_\text{eff}$ denotes an effective $RC$ time of the double cavity. It is determined by the effective conductance of the double cavity, $G_\text{eff}=G_{1\Sigma}G_{2\Sigma}/(G_{1\Sigma}+G_{2\Sigma})$ which is largest if both cavities have equal conductances.  Furthermore, it depends on the effective capacitance 
\begin{equation}
	C_\text{eff}=\frac{C_{\Sigma}(2C+C_{\Sigma})(C_{\Sigma}^2+2CC_{\Sigma}-CC_{\mu})}{2C^2C_{\mu}}
\end{equation}
describing how strong the interaction is between the two cavities. It should be minimized (without entering the Coulomb blockade regime) to maximize the rectified current.  It grows as $C^{-2}$ and for large couplings approaches the constant value $(2C_{\Sigma}-C_{\mu})C_{\Sigma}/C_{\mu}$, cf. Fig.~\ref{fig:Ceff}.  Next, as stated, the rectified current~\eqref{eq:current} is proportional to $\Lambda$, which characterizes the asymmetry of the system: 
The system is asymmetric if either the left-right conductances and/or their energy 
derivatives differ. Finally, the current~\eqref{eq:current} is linear in the applied temperature difference, so the rectified current is zero in global thermal equilibrium, as must be the case in order to satisfy the second law of thermodynamics.  We note that the sign of the current flips under either exchange of the system lead nonlinearity or under exchange of the cavity temperatures.

As the energy-dependent part of the transmission does not scale with the number of transport channels, the current Eq.~\eqref{eq:current} also turns out to be independent of the channel number. For realistic values~\cite{van_wees_quantum_1991,patel_properties_1991,roessler_transport_2011} of $C_\text{eff}=\unit[10]{fF}$, $G'=(e^2/h)\unit[]{(mV)}^{-1}$ and $\Theta_2-\Theta_1=\unit[1]{K}$, we find $I\sim\unit[0.1]{nA}$ which can be readily detected in current experiments and is two orders of magnitude larger than currents through typical Coulomb-blockaded dots.

In order to convert the heat extracted from the hot reservoir into useful work, we have to make the noise-induced current flow against a finite bias voltage $V_{1\text{L}}-V_{1\text{R}}$. The bias induces a counterflow of current given to leading order in the nonlinearity by $G_1 (V_{1\text{L}}-V_{1\text{R}})$, thus reducing the total current. At the stopping voltage, $V_\text{stop}= \Lambda\, \kB(\Theta_1-\Theta_2)/ (G_1 \tau_{RC})$, there is no current flowing through the system. The output power is given by $P=\langle I_{1\text{L}}\rangle(V_{1\text{L}}-V_{1\text{R}})$. It is parabolic as a function of the applied voltage, vanishing at zero bias and the stopping voltage; it has a maximum at half the stopping voltage given by
\begin{equation}
	P_\text{max}=\frac{\Lambda^2}{4G_1\tau_{RC}^2}(\kB(\Theta_1-\Theta_2))^2.
\end{equation}

Energy is transferred between the cavities in the form of dissipated power in the upper cavity, $P = \sum_r \langle I_{1r} V_{1r} \rangle$, the heat current given up by the lower hot cavity to the upper cold cavity, $J_H = \langle U_2 I_2 \rangle$, and the heat current given up by the upper cold cavity to its heat reservoirs, $J_C = \sum_r \langle (U_1 - V_{1r}) I_{1r} \rangle$.   It is straightforward to check that $J_H = J_C + P$ in our model, so energy is conserved in the system.  The efficiency, $\eta$, of the heat-to-charge-current converter is given by the ratio between the output power, $P$ to the inter-cavity heat current, $J_H$.
To leading order in the energy-dependent transmissions, this heat current is given by
\begin{equation}
	J_H=\frac{1}{\tau_{RC}}\kB(\Theta_2-\Theta_1),
\end{equation}
because heat will flow from hot to cold even without the nonlinearity.  The correction to this result that is linear in the voltage applied across the upper cavity is suppressed by $G'_{1r}$ (as it must to satisfy an Onsager relation, see Appendix~\ref{app:onsager}). 
As indicated earlier, the asymmetry parameter $\Lambda$ controls the process of energy-to-charge conversion, $\langle I_{1\text{L}} \rangle/J_H$.

\begin{figure}
            \includegraphics[width=\columnwidth]{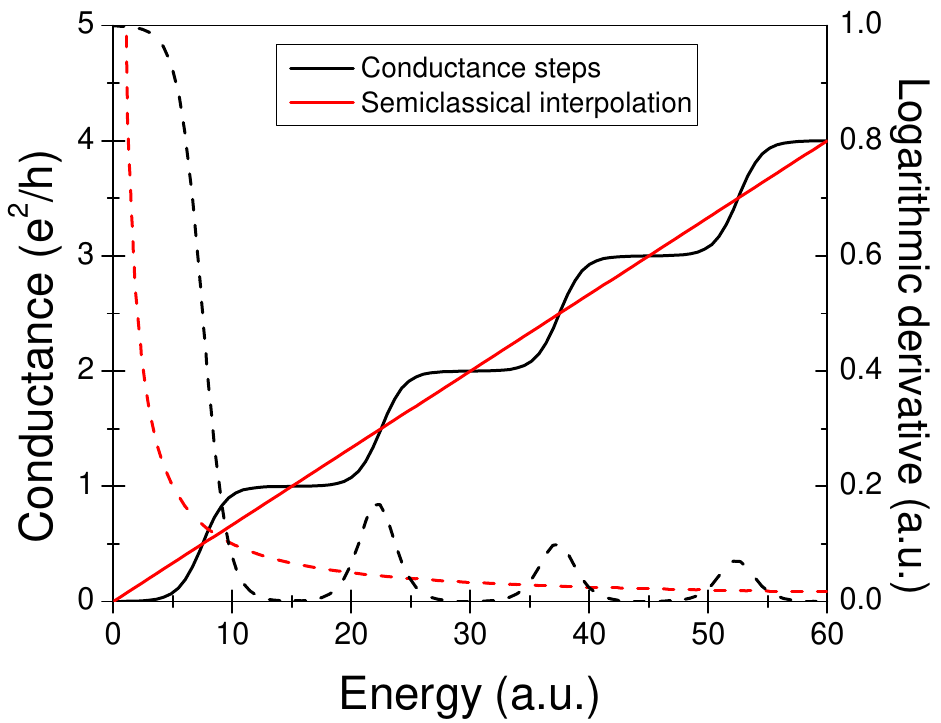}
	\caption{\label{fig:power} 
Conductance of a QPC for both fully quantized steps (full black) and semiclassical interpolation (full red). Its logarithmic derivative (black and red dashed) controls the heat-to-charge conversion, stopping voltage, power and efficiency of the energy harvester.}
\end{figure}

The efficiency $\eta=P/J_H$ exhibits the same parabolic bias dependence as the output power since the heat current is independent of the applied bias. Hence, for a given temperature difference, the maximal efficiency occurs at maximum power and is given by
\begin{equation}
	\eta_\text{max}=\frac{\Lambda^2}{4G_1\tau_{RC}}\kB(\Theta_2-\Theta_1).
\end{equation}
For $G_1=5e^2/h$ and parameters as above, we estimate $P_\text{max}\sim\unit[2]{fW}$ and a maximal efficiency of $\sim1\%$ of the Carnot efficiency for a device working at liquid-helium temperatures.
We note that while the maximal power of the system scales inversely with the number of available transport channels, the maximal efficiency even decreases with the number of channels squared. 
This is because for a large number of open channels the effect of the energy-dependence of the uppermost channel becomes less important.  This effect can be seen in Fig.~\ref{fig:power} where the logarithmic derivative of the QPC conductance is plotted, which controls the stopping voltage and other rectification figures of merit in the case where one contact is energy-independent. For a stronger nonlinearity, such as a truly step-like transmission, a nonperturbative analysis is required which could give rise to much higher efficiencies.~\cite{sokolov_energetics_1998}

In order to demonstrate that it is the nonlinearity of the rectifying cavity that is the key ingredient to our heat engine, we now briefly consider an alternative setup where a rectifier is coupled to a resistor $R$ at temperature $\Theta_2$ with a capacitor $C_R$ in parallel instead of a second cavity (see Fig.~\ref{fig:circuit} for a circuit diagram). Repeating the same analysis as above, we find that the heat-induced current, the maximal power as well as the efficiency at maximum power are given by the same expressions as above. The only difference is that the effective conductance  and capacitance take on different values, $G_\text{eff} \rightarrow G_{1\Sigma}$,  $C_\text{eff}\rightarrow [C_R C_\Sigma+C(C_R+C_\Sigma)][C_\mu C_\Sigma(C+C_\Sigma)+RG_\Sigma(CC_RC_\Sigma+(C+C_R)C_\Sigma^2-CC_RC_\mu)]/(C^2C_\mu C_\Sigma)$.

We finally apply our results to the semiclassical regime in the limit $G_{2\Sigma} \gg G_{1\Sigma}$ for simplicity. For large energies the conductance of the QPCs is given by
\begin{equation}
	G_r(E)=\frac{e^2}{h}\frac{(E-E_r)^\gamma}{\Delta_r^\gamma}\Theta(E-E_r)
\end{equation}
with $E_r$ being the energy that marks the transition from tunneling to ballistic transport and $\Delta_r$ describing how open the contact is. For equal conductances of the two QPCs at the Fermi energy, $G_\text{L}(E_F)=G_\text{R}(E_F)$, we obtain for the maximal power
\begin{equation}
	P_\text{max}=\frac{e^4}{8h}\frac{\gamma^2[\kB(\Theta_1-\Theta_2)]^2(1-\mathcal R)^2(E_F-E_\text{L})^{\gamma-2}}{\Delta_\text{L}^\gamma C_\text{eff}^2},
\end{equation}
with $\mathcal R=\Delta_\text{L}/\Delta_\text{R}$. We thus see that in order to maximize the power, we need a strong asymmetry in the contacts, $\Delta_\text{L}\ll\Delta_\text{R}$ while keeping the conductance of each contact the same. For the semiclassical result $\gamma=1/2$, we find that the power drops upon increasing the energy transport window $E_F-E_\text{L}$. We note, however, that for $\gamma>2$ the contribution from the conductance will outweigh the contribution from the nonlinearity and, thus, lead to a maximal power that increases with the transport window.  While the efficiency will drop with the inverse square of the transport window, we remark that a conductance that is exponential in energy will have an efficiency that is independent of the window size.

\section{\label{sec:conclusions}Conclusions}
We have examined a mesoscopic energy harvester consisting of a pair of quantum dots and find that as the contacts are opened, the power output can increase but typically with a drop in efficiency for the weak nonlinearity considered here. Our work demonstrates the importance of the asymmetric energy dependence of the contact transmissions. Energy harvesting from environmental fluctuations is an important goal. It might lead to nano-scale devices which can function independently of an external power supply. In densely packed electronic circuits energy harvesting might alleviate the heat removal problem. Our results are useful for future experiments that realize solid state energy harvesters. 

\acknowledgments
We acknowledge support from the project NANOPOWER (FP7/2007-2013) under Grant No. 256959.  ANJ acknowledges support from NSF Grant No. DMR-0844899 and the University of Geneva. RS was supported by  
the CSIC and FSE JAE-Doc program, the Spanish MAT2011-24331, and the ITN Grant No. 234970 (EU).

\appendix

\begin{figure}
	\includegraphics[width=\columnwidth]{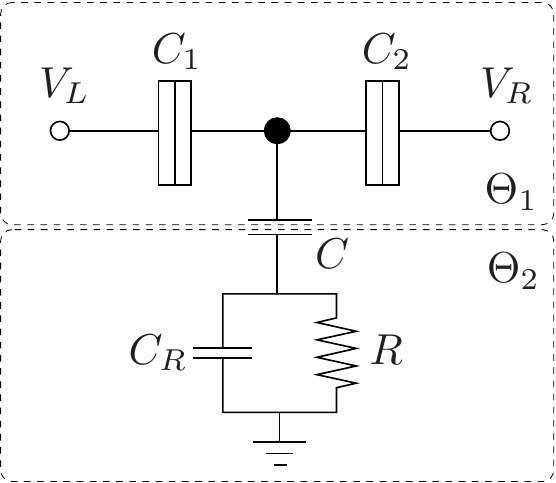}
	\caption{\label{fig:circuit}Circuit diagram of a cavity coupled to a hot resistor.}
\end{figure}

\section{\label{app:kinetic}Kinetic equation}
Our starting point is the set of kinetic equation for the distribution functions $f_i$ of the cavities,
\begin{equation}\label{eqAPP:kinetic}
	\frac{df_i}{dt}=\frac{\partial f_i}{\partial U_i}\dot U_i+\frac{1}{h\nu_{i\text{F}}}\sum_r T_{ir}(f_{ir}-f_i)+\frac{1}{e\nu_{i\text{F}}}\delta i_\Sigma.
\end{equation}

In the following, we write the distribution function as a constant part that is given by the average of the Fermi functions of the reservoirs weighted with the transmission of the respective QPC and a fluctuating part $\delta f_i$:
\begin{align}
	f_i&=\sum_r\frac{T_{ir}f_{ir}}{T_{i\Sigma}}+\delta f_i\nonumber\\
	&=\sum_r\frac{G_{ir}}{G_{i\Sigma}}f_{ir}-\Lambda_{i\text{LR}}(f_{i\text{L}}-f_{i\text{R}})\left(\delta U_i+\frac{G'_{i\Sigma}}{G_{i\Sigma}}(\delta U_i)^2\right)\nonumber\\
	&\phantom{=}+\delta f_i
\end{align}
Here, in the last step, we used $T_{ir}=T^0_{ir}-eT'_{ir} \delta U_i$ and expanded the whole expression up to second order in $\delta U_i$. We, furthermore, introduced the asymmetry parameter
\begin{equation}
	\Lambda_{i\text{LR}}=\frac{G'_{i\text{L}}G_{i\text{R}}-G'_{i\text{R}}G_{i\text{L}}}{G_{i\Sigma}^2}
\end{equation}
and abbreviated $G_{ir}=(e^2/h)T^0_{ir}$, $G'_{ir}=(e^3/h)T'_{ir}$, where $r=\text{L},\text{R},\Sigma$ refers to the left QPC, the right QPC or the sum over all QPCs adjacent to cavity $i=1,2$.

In order to relate the fluctuating part of the distribution function $\delta f_i$ to the fluctuating part of the potential $\delta U_i$, we express the charge $Q_{ic}$ inside cavity $i$ once in terms of its distribution function and once in terms of the potentials and capacitances,
\begin{align}
	Q_{ic}&=e\nu_{i\text{F}}\int dE\left(\sum_r\frac{T_{ir}}{T_{i\Sigma}}f_{ir}+\delta f_i\right)-e^2\nu_{i\text{F}}U_i,\\
	Q_{ic}&=\sum_iC_{ir}(U_i-V_{ir})+C_{ig}(U_i-V_{ig})+C(U_i-U_{\bar i}),
\end{align}
where $\bar i$ denotes the index opposite to $i$. Equating the fluctuating parts of both equations, we find
\begin{multline}\label{eqAPP:deltafdeltaU}
	\int dE\delta f_i=e\left(\frac{C_{i\Sigma}}{C_{i\mu}}+\frac{\chi_i}{e^3\nu_{i\text{F}}}\right)\delta U_i+\frac{\chi_i}{e\nu_{i\text{F}}}\frac{T'_{i\Sigma}}{T^0_{i\Sigma}}(\delta U_i)^2\\
	+e\frac{C}{e^2\nu_{i\text{F}}}(\delta U_i-\delta U_{\bar i}).
\end{multline}
Here, we introduced $\chi_i=\Lambda_{i\text{LR}}(V_{i\text{L}}-V_{i\text{R}})$, the total capacitance of cavity $i$, $C_{i\Sigma}=\sum_r C_{ir}+C_{ig}$, as well as its electrochemical capacitance $C_{i\mu}=e^2\nu_{i\text{F}}C_{i\Sigma}/(e^2\nu_{i\text{F}}+C_{i\Sigma})$.

Using Eq.~\eqref{eqAPP:deltafdeltaU} to eliminate $\delta f_i$ from the energy-integrated kinetic equation~\eqref{eqAPP:kinetic}, we obtain a set of coupled, nonlinear Langevin equations that determine $\delta U_i$,
\begin{widetext}
\begin{equation}\label{eqAPP:Langevin}
\begin{split}
	(C_{i\Sigma}+C)\dot{\delta U_i}-C\dot{\delta U}_{\bar i}=&-G_{i\Sigma}\left(\frac{C_{i\Sigma}}{C_{i\mu}}+\frac{\chi_i}{e^3\nu_{i\text{F}}}\right)\delta U_i+G'_{i\Sigma}\frac{C_{i\Sigma}}{C_{i\mu}}(\delta U_i)^2+\delta I_{i\Sigma}\\&
	-G_{i\Sigma}\frac{C}{e^2\nu_\text{iF}}(\delta U_i-\delta U_{\bar i})+G'_{i\Sigma}\frac{C}{e^2\nu_\text{iF}}\left[(\delta U_i)^2-\delta U_i\delta U_{\bar i}\right]
\end{split}
\end{equation}

\section{\label{app:diffusion}Diffusion coefficients}
The diffusion in Eq.~\eqref{eqAPP:Langevin} is characterized by the diffusion coefficients defined as~\cite{SPblanter_shot_2000}
\begin{equation}
	\langle\delta I_{ir}(t)\delta I_{ir}(0)\rangle=\frac{2e^2}{h}\int dE T_{ir}\left[f_{ir}(1-f_{ir})+f_i(1-f_i)+(1-T_{ir})(f_{ir}-f_i)^2\right]\delta(t)=D_{ir}\delta(t).
\end{equation}
\end{widetext}
Importantly, the diffusion coefficients $D_{ir}$ depend themselves on $\delta U_i$ through the energy-dependence of the transmissions $T_{ir}$. This leads to a certain ambiguity when converting the Langevin equation into a Fokker-Planck equation, see below. Evaluating the above integral and expanding the diffusion coefficient to linear order in the applied voltage, we obtain
\begin{equation}
	D_{ir}=4\kB \Theta_{i}(G_{ir}-G'_{ir}\delta U_i).
\end{equation}

\section{\label{app:FP}Fokker-Planck equation}
Given a nonlinear Langevin equation of the form
\begin{equation}\label{eqAPP:MultLangevin}
	\dot{x_i}=f_i(\vec x)+g_{ij}(\vec x)\eta_j(t)
\end{equation}
where $\vec x=(x_1,x_2,\cdots)$ and $\eta_j(t)$ is a noise source satisfying $\langle\eta_j(t)\rangle=0$ and $\langle\eta_i(t)\eta_j(t')\rangle=\delta_{ij}\delta(t-t')$, one can show~\cite{SPlau_state-dependent_2007} that it is equivalent to a Fokker-Planck equation of the form
\begin{equation}
	\frac{\partial P}{\partial t}=\frac{\partial}{\partial x_i}\left[-\left(f_i+\alpha\frac{\partial g_{il}}{\partial x_k}g_{kl}\right)P+\frac{1}{2}\frac{\partial}{\partial x_j}\left(g_{il}g_{jl}P\right)\right]
\end{equation}
where Einstein's sum convention is implied. The parameter $\alpha$ takes the values $0$ in the It\^o prescription, $1/2$ in the Stratonovich prescription and $\alpha=1$ in the Klimontovich prescription.~\cite{klimontovich_ito_1990} In our analysis, it turns out that only the Klimontovich prescription gives vanishing currents in global thermal equilibrium. In our problem, we have $g_{ij}=\delta_{ij}g_i$ such that the Fokker-Planck equation simplifies to
\begin{equation}
	\frac{\partial P}{\partial t}=\frac{\partial}{\partial x_i}\left[-\left(f_i+\alpha\frac{\partial g_i}{\partial x_i}g_i\right)P+\frac{1}{2}\frac{\partial}{\partial x_i}\left(g_i^2P\right)\right]
\end{equation}
Multiplying the Fokker-Planck equation with $x_k$ and $x_kx_l$, respectively, and integrating over all variables $x_i$, we obtain the following equations for the expectation values
\begin{align}
	\frac{d}{dt}\langle x_k\rangle&=\left\langle f_k\right\rangle+\alpha \left\langle\frac{\partial g_k}{\partial x_k}g_k\right\rangle,\\
	\frac{d}{dt}\langle x_kx_l\rangle&=\left\langle x_lf_k+x_kf_l\right\rangle+\alpha\left\langle x_l\frac{\partial g_k}{\partial x_k}g_k+x_k\frac{\partial g_l}{\partial x_l}g_l\right\rangle\nonumber\\
	&\phantom{=}+\delta_{kl}\left\langle g_k^2\right\rangle.
\end{align}
To make a closer connection to the discussion above, we introduce $g_i=\sqrt{2\tilde D_i}$ with the diffusion constants $\tilde D_i$ and obtain
\begin{align}
	\frac{d}{dt}\langle x_k\rangle&=\left\langle f_k\right\rangle+\alpha \left\langle\frac{\partial \tilde D_k}{\partial x_k}\right\rangle,\\
	\label{eqAPP:xixj}
	\frac{d}{dt}\langle x_kx_l\rangle&=\left\langle x_lf_k+x_kf_l\right\rangle+\alpha\left\langle x_l\frac{\partial \tilde D_k}{\partial x_k}+x_k\frac{\partial \tilde D_l}{\partial x_l}\right\rangle\nonumber\\
	&\phantom{=}+2\delta_{kl}\left\langle \tilde D_k\right\rangle.
\end{align}
By comparing the equation for $\langle x_k\rangle$ with the original Langevin equation, we furthermore obtain for the expectation value of the random currents
\begin{equation}\label{eqAPP:expectationdeltaI}
	\left\langle\delta I_{ir}\right\rangle=\left\langle\sqrt{2\tilde D_{ir}}\eta_{ir}(t)\right\rangle=\alpha\left\langle\frac{\partial\tilde D_{ir}}{\partial x_i}\right\rangle.
\end{equation}

For the calculation of heat current, we also need correlators of the form $\langle x_i g_{ij}(\vec x)\eta_{j}(t)\rangle$. In order to obtain them, we multiply the Langevin equation~\eqref{eqAPP:MultLangevin} by $x_i$,
\begin{equation}
	\frac{1}{2}\frac{d}{dt}x_i^2=x_if_i(\vec x)+x_ig_{ij}(\vec x)\eta_j(t).
\end{equation}
Taking expectation values and equating with Eq.~\eqref{eqAPP:xixj}, we find
\begin{equation}\label{eqAPP:potcurr}
	\langle x_i\delta I_{ir}\rangle=\langle x_i\sqrt{2\tilde D_{ir}}\eta_{ir}(t)\rangle=\frac{1}{2}\langle\tilde D_{ir}\rangle+\alpha\left\langle x_i\frac{\partial \tilde D_{ir}}{\partial x_i}\right\rangle.
\end{equation}

\section{\label{app:currents}Charge and heat currents}
The charge current between lead $r$ and dot $i$ is given by
\begin{equation}
	I_{ir}=\frac{e}{h}\int dE (T^0_{ir}-eT'_{ir}\delta U_i)(f_{ir}-f)+\delta I_{ir}
\end{equation}
which can be rewritten as
\begin{widetext}
\begin{equation}
	I_{ir}=\frac{G_{ir}G_{i\bar r}}{G_{i\Sigma}}(V_{ir}-V_{i\bar r})-\frac{G'_{ir}G_{i\bar r}}{G_{i\Sigma}}(V_{ir}-V_{i\bar r})\delta U_i-\left\{(G_{ir}-G'_{ir}\delta U_{i})\left[\left(\frac{C_{i\Sigma}}{C_{i\mu}}+\frac{C}{e^2\nu_{i\text{F}}}\right)\delta U_i-\frac{C}{e^2\nu_{i\text F}}\delta U_{\bar i}\right]\right\}+\delta I_{ir}.
\end{equation}
The expectation value of the fluctuating current part is obtained from Eq.~\eqref{eqAPP:expectationdeltaI}.

The heat current between lead $r$ and cavity $i$ is given by $J_{ir}=(U_i-V_{ir})I_{ir}$. Without an applied bias voltage, we have $J_{ir}=\delta U_i I_{ir}$ and, hence,
\begin{equation}
	\langle J_{ir}\rangle=-G_{ir}\left[\left(\frac{C_{i\Sigma}}{C_{i\mu}}+\frac{C}{e^2\nu_{i\text{F}}}\right)\langle(\delta U_i)^2\rangle^{(0)}-\frac{C}{e^2\nu_{i\text F}}\langle\delta U_i\delta U_{\bar i}\rangle^{(0)}\right]+\langle\delta U_i\delta I_{ir}\rangle^{(0)},
\end{equation}
where the potential-current correlator can be obtained using Eq.~\eqref{eqAPP:potcurr}. The superscript $(0)$ on the expectation values indicates that they have to be evaluated to zeroth order in the applied bias voltage.

The heat current up to linear order is again given by $J_{ir}=\delta U_i I_{ir}$ as both the nonfluctuating part of $U_i$ as well as the expectation value of $I_{ir}$ are of first order in the bias voltage. We find
\begin{equation}
\begin{split}
	\langle J_{ir}\rangle=&\frac{G_{ir}G_{i\bar r}}{G_{i\Sigma}}\langle\delta U_i\rangle^{(0)}(V_{ir}-V_{i\bar r})-\frac{G'_{ir}G_{i\bar r}}{G_{i\Sigma}}(V_{ir}-V_{i\bar r})\langle(\delta U_i)^2\rangle^{(0)}\\&
	-G_{ir}\left[\left(\frac{C_{i\Sigma}}{C_{i\mu}}+\frac{C}{e^2\nu_{i\text{F}}}\right)\langle(\delta U_i)^2\rangle^{(1)}-\frac{C}{e^2\nu_{i\text F}}\langle\delta U_i\delta U_{\bar i}\rangle^{(1)}\right]+\langle\delta U_i\delta I_{ir}\rangle^{(1)}.
\end{split}
\end{equation}
\end{widetext}

\section{\label{app:onsager}Onsager relations}
According to Onsager~\cite{SPonsager_reciprocal_1931,SPcasimir_onsagers_1945} the linear response coefficients of charge and heat currents as a response to bias voltage and thermal gradients are related to each other. To verify the Onsager relation for our system, we expand the charge current through cavity 1 and the heat current between the two cavities to linear order in the bias $\Delta V$ applied to cavity 1 and the temperature difference $\Delta \Theta$ between the reservoirs of the two cavities,
\begin{align}
	\langle I_{1\text{L}}\rangle&=G\Delta V+L\Delta \Theta,\\
	\langle J_{2\Sigma}\rangle&=M\Delta V+N\Delta \Theta.
\end{align}
In the main text we already found that $L=\kB\Lambda_{1\text{LR}}/\tau_{RC}$. Evaluating similarly the heat current in response to an applied bias voltage, we find $M=-\kB\Theta\Lambda_{1\text{LR}}/\tau_{RC}$ in agreement with the Onsager relation $L=-M/\Theta$.


%
\end{document}